\documentclass[pra,twocolumn,preprintnumbers,amsmath,amssymb,floatfix,showpacs]{revtex4}
\usepackage{graphicx}
\usepackage{graphics}
\usepackage{psfrag}
\usepackage{hyperref}
\usepackage{multirow}
\usepackage[sort&compress]{natbib}
\usepackage{color}
\usepackage{slashed}
\usepackage{verbatim}

\begin{document}

\hypersetup{pdftitle={Scaling relations and multicritical phenomena from Functional Renormalization}}
\title{Scaling relations and multicritical phenomena from Functional Renormalization}
\date{\today}
\author{Igor Boettcher}
\affiliation{Institute for Theoretical Physics, Heidelberg University, D-69120 Heidelberg, Germany}

\begin{abstract}
We investigate multicritical phenomena in O($N$)+O($M$)-models by means of nonperturbative renormalization group equations. This constitutes an elementary building block for the study of competing orders in a variety of physical systems. To identify possible multicritical points in phase diagrams with two ordered phases, we compute the stability of isotropic and decoupled fixed point solutions from scaling potentials of single-field models. We verify the validity of Aharony's scaling relation within the scale-dependent derivative expansion of the effective average action. We discuss implications for the analysis of multicritical phenomena with truncated flow equations. These findings are an important step towards studies of competing orders and multicritical quantum phase transitions within the framework of Functional Renormalization.
\end{abstract}


\maketitle

\section{Introduction}
The antinomy of two equally justified principles makes the essence of tragedy in literature (Hegel). In the same way, the
competition of ordering principles in complex systems adds a new enthralling element to their macroscopic phenomenology. The purest manifestation of this interplay is two either intertwining or mutually excluding ordered phases in a physical system \cite{Fisher73,PhysRevB.13.412,PhysRevB.8.4270,PhysRevLett.33.813,PhysRevLett.88.059703,Pelissetto2002549,Calabrese:2003ia,PhysRevB.67.054505,PhysRevE.78.041124,PhysRevE.88.042141,PhysRevE.90.052129}. Corresponding examples can be found everywhere in physics, including magnetic systems \cite{PhysRevLett.34.1638,PhysRevB.18.6165,PhysRevLett.95.217202,PhysRevB.22.1429,PhysRevB.24.1244}, high-temperature superconductors \cite{Zhang21021997,PhysRevB.60.13070,PhysRevB.66.094501,PhysRevB.81.235108,PhysRevB.83.155125,PhysRevB.89.121116}, graphene \cite{PhysRevLett.97.146401,PhysRevB.84.113404,PhysRevB.90.041413,Classen:2015ssa},  and dense quark matter \cite{Berges:1998rc,Strodthoff:2011tz,Fukushima:2013rx}. The theoretical description of such situations is complicated by the need to accurately resolve both universal and non-universal features of the particular phase structure.

In fact, many aspects of competing orders are already universally dictated by the theory of critical phenomena. For this consider a physical system with two distinct macroscopically ordered phases which are separated from the disordered phase by means of second order phase transitions each. The corresponding phase transitions are approached by separately fine-tuning two parameters, $g_1$ and $g_2$, which can be temperature, coupling strength, etc. One may now, for instance, want to know whether there is a second order multicritical point $(g_{1\rm c},g_{2\rm c})$ in the phase diagram where both transition lines meet, or whether there is a coexistence phase of both orders.

If a second order multicritical point exists, the system in its vicinity will be described by a critical quantum field theory for the two order parameter fields. The properties of this field theory also determine whether there can be a coexistence phase. Whereas classical multicritical phenomena are typically captured by an euclidean O($N$)+O($M$)-model, bosonic self-energy effects or the presence of gapless fermions can complicate the setting for quantum phase transitions \cite{ZinnJustinBook,SachdevBook,HerbutBook}. Furthermore, even if a multicritical point exists, the microscopic parameters of the model may not lie in the basin of attraction of this fixed point, or there may be energetic reason for the second order lines to actually become first order lines  before they meet. In these cases a coexistence phase is excluded as well.

Thus the theoretical modelling of competing orders requires both a resolution of universal properties of a given system under consideration, e.g. critical phenomena, but also non-universal ones, e.g. the effective potential to resolve first order phase transitions. Furthermore, the method should be applicable beyond the realm of classical phase transitions. A versatile tool which qualifies here is provided by the Functional Renormalization Group (FRG) \cite{Wetterich:1992yh,Morris:1993qb,Berges:2000ew,Gies:2006wv,Schaefer:2006sr,Pawlowski:2005xe,Delamotte:2007pf,Kopietz2010,RevModPhys.84.299,Braun:2011pp,Boettcher201263}. It naturally captures universal aspects and critical phenomena in the long-wavelength limit, but at the same time allows to resolve system-specific properties at intermediate scales. In the following we focus on classical multicritical phenomena in O($N$)+O($M$)-models. Those are the key building blocks for more involved setups as the bosonic two-field model always appears as a subsector of the corresponding set of beta functions.

In the vicinity of one of the second order lines described above, say the one associated to tuning $g_1$, the corresponding order parameter fluctuations are captured by an effective action or free energy functional $\Gamma$ for the order parameter field $\vec{\Phi}$, which we assume to be O($N$)-symmetric. We approximate here the effective action within the \emph{local potential approximation} including a wave-function renormalization (called LPA$^\prime$ in the following) and write
\begin{align}
 \label{Int1} \Gamma^N[\vec{\Phi}] \simeq \int \mbox{d}^dx \Bigl( -Z_{\Phi} \vec{\Phi}\cdot \nabla^2 \vec{\Phi}+ V(\Phi)\Bigr),
\end{align}
where $\vec{\Phi}$ is an $N$-component vector and $\Phi=|\vec{\Phi}|$. $V(\Phi)$ is called the effective potential. The translation invariant ansatz is also applicable in the low-energy limit for spin systems close to the phase transition. The critical point of the effective action in Eq. (\ref{Int1}) is approached by fine-tuning a parameter $\sigma \to\sigma_{\rm c}$. Close to the transition line we  have a linear relation $g_1-g_{\rm c}\propto \sigma-\sigma_{\rm c}$ which links the actual physical system to the corresponding O($N$)-model. In the same way, the vicinity of the multicritical point is captured by means of an O($N$)+O($M$)-model, which we approximate within LPA$^\prime$ according to the effective action
\begin{align}
 \nonumber \Gamma^{N,M}[\vec{\Phi},\vec{\Psi}] \simeq \int \mbox{d}^dx \Bigl(&-Z_{\Phi} \vec{\Phi}\cdot \nabla^2 \vec{\Phi} -Z_{\Psi} \vec{\Psi}\cdot \nabla^2 \vec{\Psi}\\
 \label{Int2} &+V(\Phi,\Psi)\Bigr).
\end{align}
Herein, $\vec{\Phi}$ and $\vec{\Psi}$ are $N$- and $M$-component vectors, respectively. Furthermore we restrict to the three-dimensional case in the following, $d=3$.

A reasonable candidate for a second order multicritical point of the two-field model is given by the decoupled fixed point (DFP) solution. In this case $\Gamma^{N,M}[\vec{\Phi},\vec{\Psi}]=\Gamma^N[\vec{\Phi}]+\Gamma^M[\vec{\Psi}]$, such that the critical properties are inherited from the individual single-field models. This solution represents a stable fixed point of the theory provided that mixed terms in the effective action, like $\int \mbox{d}^dx \lambda_{\Phi\Psi}\Phi^2\Psi^2$, become irrelevant in the infrared. In particular, the scaling dimension of the coupling $\lambda_{\Phi\Psi}$, denoted by $\theta_3$ in the following, is negative for a stable DFP. Aharony's exact scaling relation \cite{PhysRevLett.88.059703,PhysRevLett.51.2386} states that $\theta_3$ satisfies
\begin{align}
 \label{Int3} \theta_3^{(\rm scal.)} = \frac{1}{\nu_1} + \frac{1}{\nu_2}-d
\end{align}
at the DFP. Herein, $\nu_1$ and $\nu_2$ are the usual correlation length critical exponents of the associated O($N$)- and O($M$)-models in $d$ dimensions, respectively. The latter are well-studied in the literature \cite{Pelissetto2002549}, from what the possibility of a stable DFP can be deduced immediately. Note that the DFP solution is such that it allows for a coexistence phase of both orders \cite{PhysRevE.88.042141}.

It turns out that for the important cases $N,M\leq3$ the value of $\theta_3$ is rather small. Thus, quantitatively small errors in the $\nu_i$ or violations of the scaling relation in the two-field model might turn a stable fixed point into a seemingly unstable one. Previous studies of O($N$)+O($M$)- and O($N_1$)+O($N_2$)+O($N_3$)-models with the FRG within LPA$^\prime$ indicate a small but visible violation of the scaling relation \cite{PhysRevE.88.042141,Wetzlar,PhysRevE.90.052129}. We resolve this issue and show that it is related to a truncation-related ambiguity in defining the stability matrix of the critical theory. We demonstrate that the scaling relation is valid within LPA$^\prime$ for a commonly used regularization scheme. We argue that although the scaling relation might be violated, the correct fixed point is approached during the renormalization group flow. This is of great importance for computing the phase structure of more complicated models.

This paper is organized as follow. In Sec. \ref{SecON} we discuss critical phenomena of O($N$)-models in three dimensions by determining the associated scaling potentials. Those are then used in Sec. \ref{SecONM} to study the stability of the decoupled and the isotropic fixed point in O($N$)+O($M$)-models. In particular, we compute the critical exponents of the associated multicritical points and show that Aharony's scaling relation is valid within LPA$^\prime$. In Sec. \ref{SecAmb} we comment on an ambiguity in defining the stability matrix for the truncated system, and show how it influences both critical and multicritical phenomena. We summarize our main findings in Sec. \ref{SecCon} and give an outlook on possible extension of our approach. In App. \ref{AppRec} we derive explicit algebraic expressions for the beta functions of the single-field model which are used throughout the work. Some formulas for the two-field model which are used in Sec. \ref{SecAmb} are collected in App. \ref{AppTwo}.

\section{Scaling potentials for O(N)-models}\label{SecON}
We compute the critical scaling effective actions $\Gamma^N$ and $\Gamma^{N+M}$ in Eqs. (\ref{Int1}) and (\ref{Int2}) by means of the FRG. The latter is formulated in terms of the effective average action, $\Gamma_k$, where $k$ is a momentum scale. The effective average action interpolates between the microscopic action on large scales and $\Gamma_{k=0}=\Gamma$ \cite{Morris:1993qb,Wetterich:2001kra}. Its evolution with $k$ is given by the exact Wetterich flow equation
\begin{align}
 \label{ONa} \partial_k \Gamma_k = \frac{1}{2} \mbox{Tr} \Bigl( \frac{1}{\Gamma^{(2)}_k+R_k}\partial_k R_k\Bigr).
\end{align}
For an introduction to the method we refer to Refs. \cite{Wetterich:1992yh,Morris:1993qb,Berges:2000ew,Gies:2006wv,Schaefer:2006sr,Pawlowski:2005xe,Delamotte:2007pf,Kopietz2010,RevModPhys.84.299,Braun:2011pp,Boettcher201263}. Here we note that the second functional derivative $\Gamma_k^{(2)}$ appearing on the right hand side makes the equation highly coupled and non-linear. Furthermore, the setting requires to specify a regulator $R_k$, which regularizes the infrared properties of the functional integral. Here we employ the so-called \emph{optimized regulator} for the individual O($N$)-fields \cite{PhysRevD.64.105007,Litim:2001fd}. It is diagonal in field space and given in momentum space by
\begin{align}
 \label{ONb} R_{k}(\vec{q}^2) = Z_\Phi(k^2-\vec{q}^2)\Theta(k^2-\vec{q}^2),
\end{align}
where $\Theta$ is the step function and $\vec{q}$ is the euclidean momentum. 

We seek scaling solutions of the Wetterich equation, which describe the system at its critical point. We introduce our notation for the single-field model here, but all notions immediately generalize in a straightforward way to the two-field model. By defining the dimensionless renormalized field $ \vec{\phi} = \frac{1}{Z_\Phi^{1/2} k^{(d-2)/2}}\vec{\Phi}$, the dimensionless effective potential
\begin{align}
 \label{ONc} v_k(\phi) =k^{-d} V_k(\Phi)
\end{align}
is a function of $\phi=|\vec{\phi}|$ due to O($N$)-symmetry. The scaling solution $v(\phi)$ satisfies $\dot{v}(\phi)=0$ for all $\phi$, where the dot denotes a derivative with respect to renormalization group time $t=\log(k)$, i.e. $\dot{v}_k = k \partial_k v_k$. The scale dependence of the wave function renormalization is encoded in the anomalous dimension $\eta$. We have
\begin{align}
 \label{ONd} \eta = - \frac{1}{Z_\Phi} \dot{Z}_\Phi.
\end{align}
In the following we resolve the functional form of the effective potential $v(\phi)$, but approximate the $\phi$-dependence of $\eta$ by evaluating it at the minimum of $v(\phi)$, see Eq. (\ref{ON4}). This constitutes a truncation of the more general function $\eta(\phi)$.  With the described parametrization and regularization scheme we truncate the exact equation (\ref{ONa}), such that exact relations will typically be violated. On the other hand, this allows for an approximate solution of the flow equation.

The scaling potential for the O($N$)-model within LPA$^\prime$ from the FRG with the optimized regulator is found as the solution of
\begin{align}
 \label{ON1} 0 = -d v(\phi) + a \phi v'(\phi) + \frac{b}{1+v''(\phi)} + \frac{b(N-1)}{1+\phi^{-1}v'(\phi)},
\end{align}
where
\begin{align}
 \label{ON2} a &= \frac{d-2+\eta}{2},\ b= \frac{4v_d}{d} \Bigl(1-\frac{\eta}{d+2}\Bigr),\\
 \label{ON3} v_d &= \frac{1}{2^{d+1}\pi^{d/2}\Gamma(d/2)},
\end{align}
and the dimension is set by $d=3$. The value of the anomalous dimension $\eta$ has to be determined self-consistently such as to satisfy
\begin{align}
 \label{ON4} \eta =  \frac{8v_d}{d}\frac{\lambda^2}{\kappa^2(1+\lambda)^2},
\end{align}
where $\kappa$ is the minimum of $v(\phi)$ and $\lambda=v''(\kappa)$. The boundary conditions for the potential at the origin are $v'(0)=0$ and $v''(0)=\sigma$. Herein, $\sigma$ is a relevant parameter which has to be fine-tuned such that there exists a solution of Eq. (\ref{ON1}) for all $\phi$. Of course, $v''(0)=\sigma$ translates to $v(0)=bN/d(1+\sigma)$. One could subtract a suitable term from Eq. (\ref{ON1}) such that $v(0)=0$ since $v(0)$ has no physical significance.

Eq. (\ref{ON1}) has a trivial constant solution with $\eta=\sigma=0$, which corresponds to the Gaussian fixed point. The constant solution, however, is unstable with respect to perturbations due to the operators $\phi^2$ and $\phi^4$. The number of relevant (dangerous) perturbations of the solution decides on the likeliness of the latter to be realized in a physical system.
We seek solutions to Eq. (\ref{ON1}) which have only one relevant direction. In three dimensions, such a solution exists for all $N$. It corresponds to the Wilson--Fisher fixed point of the renormalization group trajectory. To simplify the following discussion, we refer to the ``solution'' of Eq. (\ref{ON1}) as the one solution $v_\star(\phi)$ which has only one relevant direction. The latter requires fine-tuning of $\sigma$. Further solutions might exist, but are not of interest to us here.

For large field amplitudes the solution of Eq. (\ref{ON1}) behaves as $v(\phi) \sim \phi^{d/a}$. This scaling regime, however, cannot persist to small values of $\phi$ as the term $1/(1+v''(\phi))$ breaks the invariance of $v(\phi)$ with respect to a rescaling of $\phi$. Furthermore, the minimum $\kappa$ turns out to be such that $\kappa < 0.5$ for all relevant cases. Hence it is dominated by the region of small field amplitudes, and, accordingly, also $\lambda$ and $\eta$ can be deduced from a sufficiently precise resolution of the function $v(\phi)$ for small $\phi$. This observation provides the basis for Taylor expansion schemes based on the shooting method \cite{Morris:1994ie,Morris:1994ki,Codello:2012sc,PhysRevLett.110.141601}.

For the shooting method, Eq. (\ref{ON1}) is treated as the evolution of $v(\phi)$ in the formal time-variable $\phi$ with initial conditions $v'(0)=0$ and $v''(0)=\sigma$. For a detailed introduction to the method we refer to Ref. \cite{Codello:2012sc} and restrict to a simple sketch here. Given a value for the parameter $\eta$, say $\eta=0$, there exists exactly one $\sigma=\sigma(\eta)<0$ such that Eq. (\ref{ON1}) can be integrated up to very large $\phi$, possibly $\phi\to\infty$. In practice, the right $\sigma(\eta)$ can be found to arbitrary precision by scanning possible candidates for $\sigma$ in nested intervals. The obtained potential $v(\phi,\eta)$ will typically not satisfy Eq. (\ref{ON4}) for the anomalous dimension. However, by iterating this step while using $\eta$ from the previous step, the solution collapses to the scaling solution $v_\star(\phi) = v(\phi,\eta_\star)$ rather quick in three dimensions.

\begin{figure}[t]
\centering
\includegraphics[width=8cm]{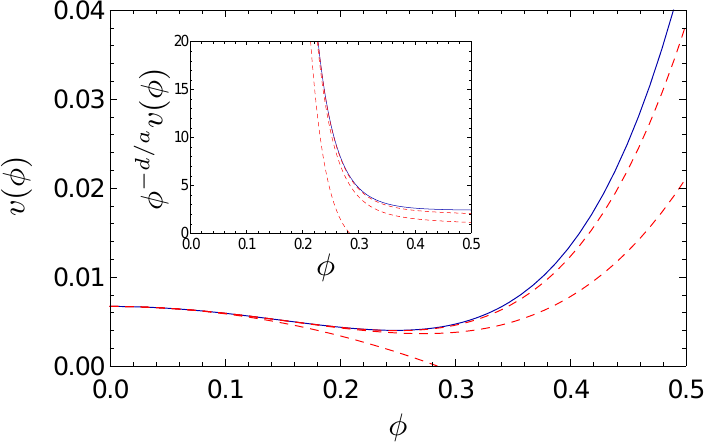}
\caption{Scaling potential for the Ising model ($N=1$) determined from Eq. (\ref{ON1}). The solid (blue) line shows $v_\star(\phi)$ obtained with the shooting method. The dashed (red) lines display the corresponding Taylor expansion with $\sigma_{\rm c}$ found from the shooting solution. We show expansions to order $\phi^2$, $\phi^4$, and $\phi^6$. The polynomial ansatz to order $\phi^8$ cannot be distinguished from the solid line within the resolution of this plot. The inset shows $\phi^{-d/a}v(\phi)$ approaching a constant value for large field amplitudes.}
\label{FigPotO1}
\end{figure}

The scaling potentials for O($N$)-models found from the shooting method are numerically exact within LPA$^\prime$. We show an example for $N=1$ in Fig. \ref{FigPotO1}. However, the important region around the minimum $\kappa$, which determines $\eta$, is also well-captured by a Taylor expansion 
\begin{align}
 \label{ON5} v(\phi) = \sum_{n=0}^L \frac{v_n}{n!} \phi^n
\end{align}
around $\phi=0$ with sufficiently large $L$. The coefficients $v_n$ can be obtained from inserting this ansatz into Eq. (\ref{ON1}) with the right $\eta$ and $\sigma_{\rm c}$ known from shooting. In fact, the coefficient $v_n$ can be expressed as an explicit polynomial in the coefficients $v_{n-1},\dots,v_0$, see Eq. (\ref{Rec8}). As a consequence, the full set of coefficients $\{v_n\}$ can recursively be determined from $v_2=\sigma_{\rm c}$, which allows for very large $L$ \cite{Morris:1994ki}. Our procedure for a given O($N$)-model is thus as follows:
\begin{itemize}
 \item[(1)] Determine $\sigma_{\rm c}$ and $\eta$ with the shooting method.
 \item[(2)] Compute the coefficients $v_n$ from the algebraic recursion relation (\ref{Rec8}) starting from $v_1=0$ and $v_2=\sigma_{\rm c}$.
\end{itemize}

The explicit recursion formula (\ref{Rec8}) for the coefficients $v_n$ allows for a fully algebraic solution of Eq. (\ref{ON1}) as well \cite{Litim:2002cf}. We name this Taylor shooting method. In this approach we determine $\sigma(\eta)$  such that the (squared) right hand side of Eq. (\ref{ON1}), after inserting the polynomial ansatz (\ref{ON5}) with a given $\sigma$, is smaller than a certain $\varepsilon>0$ for all $\phi \leq \phi_{\rm max}$. For large enough $L$, the result for $\sigma$ converges to $\sigma_{\rm c}$ as obtained from the shooting method. Accordingly, both methods yield the same Taylor expansion (\ref{ON5}) and the same critical exponents. Although the Taylor shooting method is conceptually interesting, as it is fully algebraic, the shooting method is much more efficient in finding $\sigma_{\rm c}$ for practical purposes.

The flow equation for the coefficient $v_n$ is given by $\beta_n=(\partial^n\beta/\partial\phi^n)_{\phi=0}$, where $\beta(\phi)$ is the right hand side of Eq. (\ref{ON1}). The same algebraic recursion relations which determine the set $\{v_n\}$ can be used to find $\{\beta_n\}$, see Eq. (\ref{Rec6}). We define the stability matrix $\mathcal{M}$ of the set of differential equations for $\{v_n\}$ by its entries
\begin{align}
 \label{ON6} \mathcal{M}_{nm}[v_\star] = -\frac{\partial \beta_n}{\partial v_m}\Bigr|_{v=v_\star}.
\end{align}
The derivative is applied for fixed $\eta$. Due to the overall minus sign, relevant infrared directions are signalled by positive eigenvalues of $\mathcal{M}$. We order the set of eigenvalues $\{\theta_i\}$ of $\mathcal{M}$ such that $\theta_1 > \theta_2 > \dots > \theta_L$. For sufficiently large $L$ we have
\begin{align}
  \label{ON7} \nu = \frac{1}{\theta_1},
\end{align}
where $\nu$ is the usual correlation length exponent. We display the values of $\eta \equiv \eta_\star$ and $\nu$ found from the shooting method in Tab. \ref{TableON}. For comparison we display reference values from Monte Carlo simulations and the $\varepsilon$-expansion with $\varepsilon=4-d$ in Tab. \ref{TableONref}. The Gaussian fixed point with $v(\phi)=v_0$ yields $\nu=1/2$ and $\eta=0$. In this case $\mathcal{M}$ has eigenvalues $(2,1,0,-1,-2,\dots)$.

\begin{table}
\begin{tabular}[t]{|c||c|c|c|c|}
\hline
Model & $\sigma_{\rm c}$ & $\eta$ &  $\theta_1$ & $\nu=\frac{1}{\theta_1}$ \\
\hline\hline
 O(1) & -0.16574 & 0.0443 & 1.545 & 0.647\\
 O(2) & -0.20460 & 0.0437 & 1.435 & 0.697\\
 O(3) & -0.23588 & 0.0409 & 1.347 & 0.742\\
 \hline
\end{tabular}
\caption{Critical exponents for three-dimensional O($N$)-models within LPA$^\prime$ from the FRG with the optimized regulator. The stability matrix eigenvalue $\theta_1$ and the critical exponent $\nu$ are evaluated from the stability matrix of a polynomial ansatz to order $L=30$ in Eq. (\ref{ON5}). The results are well-converged at this order. The value of $\sigma_{\rm c}$ can be  computed efficiently with the shooting method. For  demonstration purposes we show the leading digits of $\sigma_{\rm c}$, although we take into account more digits for the numerical evaluation.}
\label{TableON}
\end{table}

\begin{table}
\begin{tabular}[t]{|c||c|c|c|c|}
\hline
 \ & \multicolumn{2}{|c|}{Monte Carlo} & \multicolumn{2}{|c|}{$\varepsilon$-Expansion} \\
 \hline\hline
Model & $\nu$ & $\eta$ &  $\nu$ & $\eta$ \\
\hline
 O(1) &  0.63002(10) & 0.03627(10) & 0.6290(25) & 0.0360(50) \\
 O(2) & 0.6717(1)     &  0.0381(2)   & 0.6680(35) & 0.0380(50) \\
 O(3) & 0.7112(5)     & 0.0375(5)    & 0.7045(55) & 0.0375(45) \\
 \hline
\end{tabular}
\caption{Reference values for the critical exponents $\nu$ and $\eta$ of the O($N$)-model from Monte Carlo simulations and the $\varepsilon$-expansion with $\varepsilon=4-d$. The Monte Carlo values for $N=1, 2, 3$ are taken from Refs. \cite{PhysRevB.82.174433}, \cite{PhysRevB.74.144506}, \cite{PhysRevB.65.144520}. The results of the $\varepsilon$-expansion are from Ref. \cite{Guida:1998bx} (labelled as ``free'' therein). The deviation of the reference critical exponents from the ones obtained in this work are not related to the shooting method, but root in the simplified LPA$^\prime$-ansatz for the scaling effective action in Eq. (\ref{Int1}).}
\label{TableONref}
\end{table}

\section{Stability of fixed points for O(N)+O(M)-models}\label{SecONM}
The scaling solutions found for O($N$)-models can be used to obtain information on multicritical phenomena in O($N$)+O($M$)-models. More precisely, we can deduce the stability of the isotropic fixed point (IFP) and  decoupled fixed point (DFP) for the two-field model. The input for this consists in the values of $\sigma_{\rm c}(N)$ and $\eta(N)$ from the O($N$)-models. They allow to construct the $v_n$ of the O($N$)-model scaling solution $v_\star(\phi)$, and from this we compute the stability matrix of the O($N$)+O($M$)-model at the IFP and DFP, respectively. 

In the following we consider the two-field model along the lines of the single-field model, with an analogous notation. For the two dimensionless renormalized fields we write $\psi$ and $\chi$ to distinguish them from the field $\phi$ of the single field model. We also label them by ``1'' and ``2''.

The beta function for the effective potential $v(\psi,\chi)$ of the O($N$)+O($M$)-model from the FRG within LPA$^{\prime}$ \cite{PhysRevE.88.042141} is given by
\begin{align}
 \nonumber \beta(\psi,\chi) = &\mbox{ }-d v + a_1 \psi v_\psi + a_2 \chi v_\chi \\
 \nonumber &\mbox{ }+ \frac{b_1(1+v_{\chi\chi})+b_2(1+v_{\psi\psi})}{(1+v_{\psi\psi})(1+v_{\chi\chi})-v_{\psi\chi}^2} \\
 \label{ONM3} &\mbox{ }+ \frac{b_1(N-1)}{1+\psi^{-1}v_\psi}+\frac{b_2(M-1)}{1+\chi^{-1}v_\chi},
\end{align} 
where a subscript denotes a partial derivative with respect to the corresponding variable. The expressions for $a_{1,2}$ and $b_{1,2}$ coincide with those for $a$ and $b$ in Eq. (\ref{ON2}) when replacing $\eta \to \eta_{1,2}$.

Finding the scaling solution $v_\star(\psi,\chi)$ from $\beta(\psi,\chi)=0$ for all $\psi,\chi$ is a rather involved task. However, two  candidates which immediately come into mind are the ones associated with the IFP and the DFP. The scaling potential $v_\star(\psi,\chi)$ of the two-field model at the IFP is given by
\begin{align}
 \label{ONM1} v_\star(\psi,\chi) = v_\star^{N+M}(\phi),
\end{align}
where $\phi^2 = \psi^2 + \chi^2$ and $v_\star^{N+M}(\phi)$ is the scaling potential of the O($N+M$)-model. The system then possesses an enhanced symmetry. At the DFP, instead, we have
\begin{align}
 \label{ONM2} v_\star(\psi,\chi) = v^N_\star(\psi) + v^M_\star(\chi),
\end{align}
where $v_\star^N$ and $v_\star^M$ are the scaling solutions for the single-field O($N$)- and O($M$)-models. Both field variables are then independent from each other. At the IFP we have $\eta_1=\eta_2=\eta^{N+M}$, whereas $\eta_1=\eta^N$ and $\eta_2=\eta^M$ at the DFP.

In order to decide on the stability of a scaling potential $v_\star(\psi,\chi)$ we compute the corresponding stability matrix $\mathcal{M}$ of the two-field model. A stable fixed point has only two relevant directions and thus there are exactly two positive eigenvalues of $\mathcal{M}$. The sign of the third eigenvalue, $\theta_3$, decides on the stability of $v_\star(\psi,\chi)$. To compute the stability matrix, we insert the formal ansatz 
\begin{align}
 \label{ONM4} v(\psi,\chi) = \sum_{n=0}^L \sum_{m=0}^{L-n} \frac{v_{nm}}{n!m!} \psi^n \chi^m
\end{align}
into Eq. (\ref{ONM3}) to obtain the beta function for the coefficient $v_{nm}$ as
\begin{align}
 \label{ONM5} \beta_{nm} = \frac{\partial^n}{\partial \psi^n} \frac{\partial^m}{\partial \chi^m} \beta(\psi,\chi) \Bigr|_{\psi=\chi=0}.
\end{align}
The entries of the stability matrix associated to the solution $v_\star(\psi,\chi)$ are given by
\begin{align}
 \label{ONM6} \mathcal{M}_{nm,n'm'}[v_\star] = - \frac{\partial\beta_{nm}}{\partial v_ {n'm'}}\Bigr|_{v=v_\star}.
\end{align}
As in the single-field case, we keep $\eta_1$ and $\eta_2$ fixed when applying the derivative.

We write $v_\star(\phi)=\sum_n \frac{v_n^N}{n!}\phi^n$ for the scaling solution of the single-field O($N$)-model. All $v_n$ with odd $n$ vanish. The scaling potential at the IFP with $N=M$ then has coefficients
\begin{align}
 \label{ONM7} v_{2n,2m} = \frac{(2n)!(2m)!(n+m)!}{(2n+2m)!n!m!} v^{2N}_{2n+2m}.
\end{align}
At the DFP we have
\begin{align}
 \label{ONM8} v_{nm} = v^N_n \delta_{m0} +v^M_m\delta_{n0}
\end{align}
with Kronecker delta $\delta_{nm}$. 

We summarize the results for the eigenvalues of the stability matrix at the IFP and DFP in Tabs. \ref{TableIFP} and \ref{TableDFP}. We confirm the expectation that $\theta_2$ at the IFP and $\theta_1$ and $\theta_2$ at the DFP can be deduced from the knowledge of the eigenvalues of the critical single-field models. (The same is true for further irrelevant eigenvalues, which we do not display here, see for instance  \cite{PhysRevE.88.042141}.) The remaining eigenvalues, however, are genuinely determined from the two-field model, and are thus a consequence of the approximations which lead to Eq. (\ref{ONM3}) for the beta function of the O($N$)+O($M$)-model.

\begin{table}
\begin{tabular}[t]{|c||c|c|c|}
\hline
Model  & $\theta_1$ & $\theta_2$ & $\theta_3$  \\
\hline\hline
 O(1)+O(1) & 1.756 & \emph{1.435} & -0.042\\
 O(1.15)+O(1.15)& 1.767 & \emph{1.406} & -0.001\\
 O(1.25)+O(1.25) & 1.774 & \emph{1.388} & 0.025 \\
 O(1.5)+O(1.5) & 1.790 & \emph{1.347} & 0.086 \\ 
 \hline
\end{tabular}
\caption{Stability of the IFP for the three-dimensional O($N$)+O($M$)-model. As the associated scaling potential only depends on the solution of the single field O($N+M$)-model, we can set $N=M$ for simplicity. We display the three largest eigenvalues of the stability matrix of the two-field model at the IFP. The second one (in italics) coincides with $\theta_1$ from the O($N+M$)-model. Within our approximation the IFP gets unstable at a critical field value of $(N+M)_{\rm c}\simeq 2.3$, where $\theta_3$ becomes positive.}
\label{TableIFP}
\end{table}

\begin{table}
\begin{tabular}[t]{|c||c|c|c||c|}
\hline
Model &  $\theta_1$ & $\theta_2$ & $\theta_3$ & $\theta_3^{(\rm scal.)}$ \\
\hline\hline
 O(1)+O(1) & \emph{1.545} & \emph{1.545} & 0.090 & 0.090\\
 O(1)+O(2) & \emph{1.545} & \emph{1.435} & -0.020 & -0.020\\
 O(1)+O(3) &  \emph{1.545} & \emph{1.347} & -0.108 & -0.108\\
 \hline
\end{tabular}
\caption{Stability of the DFP for the three-dimensional O($N$)+O($M$)-model. The eigenvalues $\theta_i$ are derived from the stability matrix of the two-field model at the DFP. The first two of them (in italics) coincide with the largest eigenvalue of the corresponding O($N$)- and O($M$)-model, see Tab. \ref{TableON}. It is a nontrivial finding that $\theta_3$ coincides with $\theta_3^{(\rm scal.)}=\theta_1+\theta_2-3$ from the scaling relation (\ref{Int3}). This shows the consistent treatment of critical and multicritical phenomena with the FRG within LPA$^\prime$. The scaling relation was found to be valid for all cases which have been tested, also including non-integer $N$ and $M$. }
\label{TableDFP}
\end{table}

In the case of the IFP the enhanced symmetry corresponding to the O($N+M$)-group allows to choose $N=M$ without loss of generality. We find the critical number of field components for the stability of the IFP to be given by $(N+M)_{\rm c}\approx 2.3$ in agreement with the FRG-study in Ref. \cite{PhysRevE.88.042141}. Using an exponential regulator function, the upper boundary has been found to be $(N+M)_{\rm c}=3.1$  with the FRG \cite{PhysRevB.65.140402}. This may be compared to $(N+M)_{\rm c}=2.89(4)$ and $(N+M)_{\rm c}=2.87(5)$ from resummed six-loop perturbation theory and constrained five-loop $\epsilon$-expansion \cite{PhysRevB.61.15136}. These results strongly suggest that the O(1)+O(1)-model has a stable IFP. If  $N+M$ slightly exceeds the critical value, the stable fixed point is given by the biconical fixed point (BFP) instead \cite{PhysRevB.67.054505,PhysRevE.88.042141}. The BFP supports a coexistence phase and is linked to tetracritical behavior. The corresponding critical exponents at the multicritical point differ from those of the IFP and DFP.

The stability region of the DFP in the $(N,M$)-plane is bounded from below by the sign change of the third eigenvalue, $\theta_3$, of the stability matrix of the two-field model. We find that $\theta_3$ at the DFP indeed agrees with the prediction for $\theta_3^{(\rm scal.)}$ from the scaling relation in Eq. (\ref{Int3}). Here we say that the scaling relation is satisfied if it is true for the  number of significant digits displayed in Tab. \ref{TableDFP}. This accuracy is probably sufficient for deciding on the stability of fixed points in all relevant physical cases. Indeed, from the results presented in Tab. \ref{TableDFP} it appears to be very unlikely that the fixed point of  an actual physical system is such that $|\theta_3|<0.001$. Although the present analysis indicates that $\theta_3=\theta_3^{(\rm scal.)}$ is valid to higher accuracy, an extensive numerical analysis of the associated convergence is outside of the scope of this work.

Within LPA$^\prime$ we find that O($N$)+O($M$)-models with integer values of $N+M>2$ support a stable DFP. Due to the validity of the scaling relation, this information can also be deduced from the critical exponents $\nu_i$ in Tab. \ref{TableON}. The slight deviation of the $\nu_i$ from the world's best values \cite{Pelissetto2002549} due to our approximation can have profound implications on the (apparent) stability of the DFP in a given model. In particular, for the O($1$)+O($2$)-model we find the DFP to be the stable fixed point. In contrast, from the Monte Carlo reference values in Tab. \ref{TableONref} we find $\theta_3^{(\rm scal.)}\simeq0.08$ for the O(1)+O(2)-case, hence suggesting an unstable DFP. For a discussion of the current status of understanding of the stable fixed point in this model we refer to Ref. \cite{PhysRevE.88.042141}.

Whereas the scaling potentials of the IFP and DFP are constructed from the solutions of the single-field O($N$)-models, the determination of scaling potentials which are genuinely multicritical, such as the one related to the BFP, requires to solve the partial differential equation $\beta(\psi,\chi)=0$ for $v(\psi,\chi)$ in both variables,  $\psi$ and $\chi$. However, inserting the Taylor expansion (\ref{ONM4}) into this differential equation does not yield a recursion relation for the $v_{nm}$. Hence the Taylor shooting method cannot be applied in a straightforward fashion. This also suggests a failure of the shooting method. The generalization of the shooting method to find generic scaling potentials $v_\star(\psi,\chi)$ in both field variables constitutes an interesting and important direction for future works on multicritical phenomena with the FRG.

\section{Truncated stability matrices}\label{SecAmb}

In this section we discuss a truncation-related ambiguity in the definition of the stability matrix $\mathcal{M}$, which leads to an apparent violation of the scaling relation (\ref{Int3}) in one case. We further show that this violation of the scaling relation does not conflict predictions on the stability of the IFP and DFP since in both cases $\theta_3$ turns out to be independent of the corresponding definition.

For this we recall that in  Eq. (\ref{ON6}) we defined $\mathcal{M}_{nm}=-\partial \beta_n/\partial v_m$ by means of a derivative where the anomalous dimension $\eta$ is kept fixed. However, due to Eq. (\ref{ON4}), $\eta$ depends on $\kappa$ and $\lambda=v''(\kappa)$. With this, $\eta$ implicitly depends on the coefficients $v_n$, and the $v_n$-derivative may be applied to $\eta(\kappa,\lambda)$ appearing in $\beta_n$ as well. In fact, both approaches for computing the stability matrix are applied in the literature.

To make the following discussion more transparent we introduce $u(\rho)=v(\sqrt{2\rho})$ and the expansion
\begin{align}
 \label{Amb1} u(\rho) = \sum_{n=2}^{K} \frac{u_n}{n!} (\rho-\rho_0)^n,
\end{align}
where $\rho=\phi^2/2$ and $\rho_0=\kappa^2/2$ such that $u'(\rho_0)=0$. The running couplings of the system are 
\begin{align}
\{g_n\}=(\rho_0,u_2,\dots,u_{K}).
\end{align}
The corresponding flow equations are given by  $\dot{\rho}_0 = -\beta'(\rho_0)/u_2,\ \dot{u}_n = \beta^{(n)}(\rho_0)+u_{n+1}\dot{\rho}_0$. Here $\beta(\rho)$ is the right hand side of Eq. (\ref{ON1}) expressed in terms of $u$. We have
\begin{align}
 \nonumber \beta(\rho) = & -d u(\rho) +2a \rho u'(\rho)+\frac{b}{1+u'(\rho)+2\rho u''(\rho)}\\
 \label{Amb3} &+\frac{b(N-1)}{1+u'(\rho)}.
\end{align}
The term $u_{n+1} \dot{\rho}_0$ originates from the implicit dependence of $u_n$ on the expansion point $\rho_0$. The anomalous dimension (\ref{ON4}) in this parametrization reads
\begin{align}
 \label{Amb4} \eta = \frac{16 v_d}{d} \frac{\rho_0 u_2^2}{(1+2u_2\rho_0)^2}.
\end{align}
The coefficients $u_n$ of the scaling solution $u_\star(\rho)$ are easily derived from the $v_n$ of $v_\star(\phi)$. The expressions for the beta functions of the O($N$)+O($M$)-model in terms of $u$ are presented in App. \ref{AppTwo}.

Our goal is to compare results for critical exponents from two different definitions of the stability matrix for O($N$)- and O($N$)+O($M$)-models, which we denote by (A) and (B) in the following. In the single-field case we either compute the stability matrix for fixed anomalous dimension, i.e.
\begin{align}
\label{AmbA} (A):\ \mathcal{M}_{nm} = -\Bigl(\frac{\partial \beta_n}{\partial g_m}\Bigr)_\eta,
\end{align}
or we account for the implicit running coupling dependence of $\eta$ via
\begin{align}
\label{AmbB} (B):\ \mathcal{M}_{nm} = -\Bigl(\frac{\partial \beta_n}{\partial g_m}\Bigr)_\eta - \Bigl(\frac{\partial \beta_n}{\partial \eta}\Bigr)_{u}\ \frac{\partial \eta}{\partial g_m}.
\end{align}
Both expressions are evaluated for $u_\star$. In the last line, the second term yields a nonvanishing contribution for $g_m=\rho_0$ or $g_m=u_2$. In the previous sections we have applied scheme (A). The generalizations of both definitions to the two-field model are given in Eqs. (\ref{AmbATwo}) and (\ref{AmbBTwo}).

We find the correlation length exponents $\nu_i$ obtained from (A) and (B) to deviate at the percent level, see Tab. \ref{TableAmb}. As a consequence, we obtain different results for $\theta_3^{(\rm scal.)}$ from Eq. (\ref{Int3}) in O($N$)+O($M$)-models at the DFP. Most dramatically, applying the scaling relation yields contradictory results on the stability of the DFP for the O(1)+O(2)-model.

\begin{table}
\begin{tabular}[t]{|c||c|c||c|c||c|c|}
\hline
Model & \multicolumn{2}{|c||}{LPA} & \multicolumn{2}{|c||}{LPA$^\prime$, (A)} & \multicolumn{2}{|c|}{LPA$^\prime$, (B)} \\
\hline\hline
  & $\eta$ & $\nu$ & $\eta$ & $\nu$ & $\eta$ & $\nu$ \\
\hline
 O(1) & 0 & 0.650 & 0.0443 & 0.647 & 0.0443 & 0.637\\
 O(2) & 0 & 0.708 & 0.0437 & 0.697 & 0.0437 & 0.686\\
 O(3) & 0 & 0.761 & 0.0409 & 0.742 & 0.0409 & 0.732\\
 \hline\hline
 DFP &  $\theta_3$ & $\theta_3^{(\rm scal.)}$ & $\theta_3$ & $\theta_3^{(\rm scal.)}$  & $\theta_3$ & $\theta_3^{(\rm scal.)}$  \\
 \hline
 O(1)+O(1) & 0.079 & 0.079 & 0.090 & 0.090 & 0.090 & 0.141 \\
 O(1)+O(2) & -0.048 & -0.048 & -0.020 & -0.020 & -0.020 & 0.028 \\
 O(1)+O(3) & -0.147 & -0.147 & -0.108 & -0.108 & -0.108 & -0.063\\
\hline
\end{tabular}
\caption{We compare critical exponents computed from truncated stability matrices where the variation of the anomalous dimension is either neglected or respected, labelled (A) and (B). They correspond to Eqs. (\ref{AmbA}),(\ref{AmbATwo}) and (\ref{AmbB}),(\ref{AmbBTwo}), respectively. Within scheme (A), which has been applied in the previous sections, and which also includes LPA as a special case, the scaling relation $\theta_3=\theta_1+\theta_2-d$ is satisfied at the DFP. In contrast, the scaling relation is violated when applying (B). However, the eigenvalue $\theta_3$ found from the two-field model coincides in both cases.}
\label{TableAmb}
\end{table}

In the two-field model, the ambiguity in defining $\mathcal{M}$ seems to only afflict those eigenvalues which are inherited from the single-field models. Those have been highlighted in italics in Tabs. \ref{TableIFP} and \ref{TableDFP}. Indeed, at the IFP, the results for $\theta_1$ and $\theta_3$ computed with (A) agree with the results of Ref. \cite{PhysRevE.88.042141} computed with (B). In contrast, the value of $\theta_2$, which is identical to $1/\nu$ of the corresponding O($N+M$)-model, disagrees due to the difference in the correlation length exponents.

A similar behavior is found at the DFP. We find the first two eigenvalues $\theta_1$ and $\theta_2$ within (A) and (B) to coincide with $1/\nu_i$ from the individual single-field models within (A) and (B), respectively. For (A), this is shown in Tab. \ref{TableDFP}. The third exponent at the DFP, $\theta_3$, is found to be independent of the prescription for computing the stability matrix. Consequently, the scaling relation $\theta_3=\theta_1+\theta_2-d$  is satisfied for (A), whereas it is violated for (B). Within a truncation which neglects the anomalous dimension, referred to as LPA, the scaling relation is also satisfied. These findings are summarized in Tab. \ref{TableAmb}.

In the analysis put forward so far we focussed on solving fixed point equations. The corresponding set of equations arises in the $k\to0$ limit of the renormalization group flow of $\Gamma_k$ for a system tuned to criticality. During the evolution with $k>0$, the apparent ambiguity between (A) and (B) is lifted. The running of couplings \emph{does} resolve the $\kappa$- and $\lambda$-dependence of the anomalous dimension. Variations in $\kappa_k$ and $\lambda_k$ will influence $\eta_k=\eta(\kappa_k,\lambda_k)$ and might drive the system away from criticality. Hence the scaling exponents of eigenperturbations which appear during the flow are determined by scheme (B). Accordingly, the scaling relation is violated within LPA$^\prime$ during the flow for a Taylor expansion of the effective potential.

It is an important finding of our investigation that the scaling dimension $\theta_3$ of the operator $\Phi^2\Psi^2$ at the IFP and DFP is independent of the definition of the stability matrix. As a result, when considering a particular physical system by means of the flow equation (\ref{ONa}), the underlying O($N$)+O($M$)-model for low $k$ resolves the stability of the IFP or DFP (i.e. the value of $\theta_3$) in a unique fashion. The scaling relation (\ref{Int3}) allows to \emph{predict} the value of $\theta_3$ at the DFP by inserting the $\nu_i$ of the O($N$)-model as obtained from scheme (A).

\section{Conclusions and outlook}\label{SecCon}

In this work we have investigated scaling solutions for O($N$)- and O($N$)+O($M$)-models in three dimensions within the framework of Functional Renormalization. The results have been obtained for a specific, but commonly applied truncation and regularization scheme for the effective average action. Our main findings are summarized in the following list.
\begin{itemize}
 \item[(1)] By a combination of shooting and algebraic recursion techniques it is possible to efficiently determine scaling solutions for O($N$)-models and to decide on the stability of the IFP and DFP for multicritical O($N$)+O($M$)-models.
 \item[(2)] Aharony's scaling relation (\ref{Int3}) is valid at the DFP within LPA$^\prime$. In particular, the value of $\theta_3$ at the DFP within our truncation can be deduced from critical exponents of O($N$)-models.
 \item[(3)] Previously found violations of the scaling relation are related to an ambiguity in defining the stability matrix of the truncated system. The scaling relation is violated during  the renormalization group flow of running couplings.
 \item[(4)] The value of $\theta_3$ at both the IFP and DFP is not affected by this ambiguity. Therefore, the stability of multicritical points is faithfully captured by the running of couplings. The violation of the scaling relation during the flow is thus reduced to a mere little blemish.
\end{itemize}

We conclude that Functional Renormalization provides a consistent picture of both critical and multicritical phenomena for scalar theories. In particular, the present truncation scheme with a Taylor expansion of the effective potential is simple enough to be applied to more complicated physical systems. We do not expect the regularization scheme to qualitatively change any of the above statements.

To reach higher quantitative precision for critical exponents in O($N$)- and O($N$)+O($M$)-models, further improvements of the truncation are required. For one, the Taylor expansion of $v(\phi)$ is expected to have a finite radius of convergence. It thus fails to resolve the asymptotic scaling behavior for large $\phi$. Although the latter is captured by the shooting method, this information is eventually lost when defining the stability matrix in terms of the associated Taylor coefficients $v_n$. A possible way around this problem is to use spectral methods \cite{Borchardt:2015rxa}, such as projection onto a complete set of orthogonal polynomials, to resolve the full functional form of $v(\phi)$. The corresponding techniques can also be used to determine the scaling dimension of eigenperturbations $\delta v(\phi)$ of the system. It will be interesting to study whether approximation schemes beyond a Taylor expansion fulfill the scaling relation also within (B), and thus eliminate the above-mentioned blemish.

Another direction of improving the present truncation consists in the inclusion of a field-dependent wave function renormalization, or kinetial, $Z_\Phi(\Phi)$. Among other changes on the right hand side of Eq. (\ref{ON1}), we will have a field-dependent anomalous dimension $\eta(\phi)$. The scaling solution then consists of two functions, $v_\star(\phi)$ and $\eta_\star(\phi)$. The corresponding field-dependence can now be computed with the same methods as introduced above. In particular, we may hope to resolve the ambiguity between (A) and (B) in this way. Indeed, it probably originates from the rather crude approximation $\eta(\phi) \approx \eta(\kappa)$ for all $\phi$. The remaining dependence of $\eta(\kappa)$ onto the running couplings is very limited in its ways to react onto perturbations. Therefore, defining the stability matrix according to (B) need not be a consistent improvement of the truncation, and (A) is the safer choice, as it can be seen as an expansion in $\eta\ll1$. 
Exciting application of the extension by means of $\eta(\phi)$ are found in lower-dimensional systems \cite{PhysRevLett.75.378,PhysRevB.64.054513,PhysRevE.90.062105}.

The results of this work provide a solid basis for studies of competing ordering phenomena in the realm of fermionic quantum phase transitions with the FRG. Therein, the beta functions for the O($N$)+O($M$)-model appear as a subset of the larger set of flow equations for the whole system. For instance, such a system could be given by the Gross--Neveu--Yukawa-model with O(1)- and O(3)-symmetric order parameter fields in 2+1 dimensions, describing multicriticality of gapless Dirac fermions  in graphene \cite{Classen:2015ssa}. Within LPA$^\prime$ it will be exciting to see whether associated scaling relations are still satisfied within (A). Furthermore, our finding of the scheme-independence of $\theta_3$ for both the IFP and DFP -- if it persists in the presence of fermions -- allows to unambiguously resolve the corresponding value of $\theta_3$ and thus the stability of multicritical points from the flow. Note that fermion-boson-couplings are in general also $\phi$-dependent, e.g. given by a Yukawa coupling $h(\phi)$, which introduces new challenges for solving the scaling equations \cite{Pawlowski:2014zaa,Vacca:2015nta}. Again, Taylor expansion schemes or spectral methods qualify as candidates to address such questions.

\begin{center}
 \textbf{Acknowledgements}
\end{center}

The author thanks M. M. Scherer, C. Wetterich, and S. Wetzel for inspiring discussions.  This work is supported by the Graduate Academy Heidelberg and ERC Advanced Grant No. 290623.

\appendix

\section{Recursion relation for expansion coefficients}\label{AppRec}

The coefficients $v_n$ for the single-field models can be derived from recursive relations. To obtain the expressions we employ Faa di Bruno's formula
\begin{align}
 \nonumber &\frac{\partial^n}{\partial x^n} f(g(x)) \\
 \label{Rec2}&= \sum_{k=0}^n f^{(k)}(g(x)) B_{n,k}(g^{(1)}(x),\dots,g^{(n-1+k)}(x)).
\end{align}
Here $B_{n,k}(X_1,\dots,X_{n-k+1})$ are the Bell polynomial defined by
\begin{align}
 \label{Rec3} B_{n,k}(X_1,\dots,X_{n-k+1}) = n! \sum_{\{m_j\}} \prod_{j=1}^{n-k+1} \frac{1}{m_j!} \Bigl(\frac{X_j}{j!}\Bigr)^{m_j},
\end{align}
with the sums being over sets $\{m_j\}=(m_1,\dots,m_{n-k+1})$ such that $\sum_j m_j=k$ and $\sum_j j m_j=n$. This combinatoric definition, however, is not of practical relevance since the Bell polynomials are implemented in many computer algebra system. For instance, in Mathematica 8 \cite{Mathematica} they can be accessed via
\begin{align}
 \label{Rec4} B_{n,k}(X_1,\dots,X_{n-k+1}) = \text{BellY}[n,k,\{X_j\}].
\end{align}
The number of elements in the set $\{X_j\} = \{X_1,\dots,X_n\}$ can be larger than $n-k$ since only the first $n-k$ elements are used for the evaluation of $B_{n,k}$. 
We set $B_{0,0}(X)=1$.

We now derive the Taylor coefficients for the O($N$)-model. The recursion is initialized by $v_0=Nb/d(1+\sigma)$ and $v_2=\sigma$. All coefficients with odd $n$ vanish. The right hand side of Eq. (\ref{ON1}) is given by
\begin{align}
 \nonumber \beta(\phi) = &-d v(\phi) + a \phi v'(\phi) + bf(v''(\phi)) \\
 \label{Rec5} &\mbox{ }+ (N-1)bf(q(\phi)),
\end{align}
where $f(x)=(1+x)^{-1}$ and $q(\phi)=\phi^{-1}v'(\phi)$. With the ansatz (\ref{ON5}) for $v(\phi)$, the expansion coefficients of $v''(\phi)$ and $q(\phi)$ at $\phi=0$ are given by $v_{n+2}$ and $v_{n+2}/(n+1)$, respectively. With Eq. (\ref{Rec2}) we find that $\beta_n=\beta^{(n)}(0)$ is given by
\begin{align}
 \nonumber \beta_n = \mbox{ }& (an-d) v_n + b\sum_{k=0}^n f_k B_{n,k}(v_3,\dots,v_{n+2}) \\
 \label{Rec6} &\mbox{ }+ (N-1) b \sum_{k=0}^n f_k B_{n,k}\Bigl(\frac{v_3}{2},\dots,\frac{v_{n+2}}{n+1}\Bigr)
\end{align}
with
\begin{align}
 \label{Rec7} f_k = \frac{(-1)^k k!}{(1+v_2)^{k+1}}.
\end{align}
The coefficients $v_n$ of the scaling solution are now found from $\beta_n = \dot{v}_n =0$. We note that Eq. (\ref{Rec6}) is linear in the highest coefficient $v_{n+2}$. This is a result of the fact that $B_{n,k}(X_1,\dots,X_{n-k+1})$ can only depend on $X_n$ for $k=1$. In this case, however, we find from Eq. (\ref{Rec3}) that the polynomial equals $X_n$. Consequently, $v_{n+2}$ is found from
\begin{align}
 \nonumber 0 = &\mbox{ }(an-d) v_n +Nb f_0 -\frac{b v_{n+2}}{(1+\sigma)^2}\Bigl(1+\frac{N-1}{n+1}\Bigr) \\
 \nonumber &\mbox{ }+ b\sum_{k=2}^n  f_k \Bigl[ B_{n,k}(v_3,\dots,v_{n+1})\\
 \label{Rec8} &\mbox{ }+ (N-1)B_{n,k}\Bigl(\frac{v_3}{2},\dots,\frac{v_{n+1}}{n}\Bigr)\Bigr].
\end{align}

\section{Flow equations for the two-field model}\label{AppTwo}
We recall the flow equations for the effective potential and the anomalous dimensions of the O($N$)+O($M$)-model within LPA$^\prime$ for the optimized regulator \cite{PhysRevB.83.155125,PhysRevE.88.042141}. We write
\begin{align}
 \label{Two1} \rho = \psi^2/2,\ \mu=\chi^2/2.
\end{align}
The dimensionless effective potential  $u$ is defined via
\begin{align}
 \label{Two2} u(\rho,\mu) = v(\psi,\chi).
\end{align}
The beta function of $u$ is given by
\begin{align}
 \nonumber \beta(\rho,\mu) = &-d u + 2 a_1 \rho u_\rho + 2 a_2 \mu u_\mu\\
 \nonumber &+\frac{b_1(1+u_\mu+2\mu u_{\mu\mu})+b_2(1+u_\rho + 2\rho u_{\rho\rho})}{(1+u_\rho + 2\rho u_{\rho\rho})(1+u_\mu+2\mu u_{\mu\mu})-4\rho\mu u_{\rho\mu}^2}\\
 \label{Two3} &+\frac{b_1(N-1)}{1+u_\rho} + \frac{b_2(M-1)}{1+u_\mu},
\end{align}
such that $\dot{u}(\rho,\mu)=\beta(\rho,\mu)$.

To compute the stability matrix of the two field model we apply the expansion
\begin{align}
 \label{Two4} u(\rho,\mu) = \sum_{n=0}^{K} \sum_{m=0}^{K-n} \frac{u_{nm}}{n!m!} (\rho-\rho_0)^n(\mu-\mu_0)^m,
\end{align}
where $(\rho_0,\mu_0)$ is the location of the minimum of the effective potential. Its scale dependence is given by
\begin{align}
 \label{Two5} \dot{\rho}_0 &=- \frac{u_{02} \beta^{(1,0)}(\rho_0,\mu_0)-u_{11}\beta^{(0,1)}(\rho_0,\mu_0)}{u_{20}u_{02}-u_{11}^2},\\
 \label{Two6} \dot{\mu}_0 &=- \frac{u_{20} \beta^{(0,1)}(\rho_0,\mu_0)-u_{11}\beta^{(1,0)}(\rho_0,\mu_0)}{u_{20}u_{02}-u_{11}^2}.
\end{align} 
The flow equations for the coefficients $u_{nm}$ with $n,m\geq 2$ read
\begin{align}
 \label{Two7} \dot{u}_{nm} = \beta^{(n,m)}(\rho_0,\mu_0) + u_{n+1,m}\dot{\rho}_0+u_{n,m+1}\dot{\mu}_0.
\end{align}
Again, the implicit dependence on the location of the minimum has been taken into account. The set of running couplings is given by
\begin{align}
\{g_{nm}\}=(\rho_0,\mu_0,\{u_{nm}\}_{n+m>1}).
\end{align}
The elements of the  stability matrix read
\begin{align}
 \label{AmbATwo} (A):\ \mathcal{M}_{nm,n'm'}=\ &-\Bigl(\frac{\partial\dot{u}_{nm}}{\partial g_{n'm'}}\Bigr)_{\eta_1,\eta_2},\\
  \nonumber (B):\ \mathcal{M}_{nm,n'm'}=\ &-\Bigl(\frac{\partial\dot{u}_{nm}}{\partial g_{n'm'}}\Bigr)_{\eta_1,\eta_2}\\
 \label{AmbBTwo}&- \sum_{i=1,2} \Bigl(\frac{\partial\dot{u}_{nm}}{\partial \eta_i}\Bigr)_{u}\ \frac{\partial \eta_i}{\partial u_{n'm'}}.
\end{align}
for the schemes (A) and (B), respectively. The expressions are evaluated for $u_\star$. This generalizes Eqs. (\ref{AmbA}) and (\ref{AmbB}) to the case of two order parameter fields.

The anomalous dimensions for the two field model read
\begin{align}
  \label{Two8} \eta_1 &= \frac{16v_d}{d} \frac{\mu_0 u_{11}^2+\rho_0(u_{20}+2\mu_0\Delta)^2}{(1+2\rho_0u_{20}+2\mu_0u_{02}+4\rho_0\mu_0\Delta)^2},\\
 \label{Two9} \eta_2 &=\frac{16v_d}{d} \frac{\rho_0 u_{11}^2+\mu_0(u_{02}+2\rho_0\Delta)^2}{(1+2\rho_0u_{20}+2\mu_0u_{02}+4\rho_0\mu_0\Delta)^2}
\end{align} 
with $\Delta=u_{20}u_{02}-u_{11}^2$. At the DFP we have $u_{\star,11}=0$ and $\Delta_\star=u_{\star,20}u_{\star,02}$, and the expressions generalize to the anomalous dimensions of the individual single field models.


\bibliographystyle{apsrev4-1}
\bibliography{refs_scaling_pots} 

\end{document}